\input{aipcheck}

\documentclass[
    ,final            
  ]
  {aipproc}

\layoutstyle{6x9}

\begin{document}

\title{Where lies the Peak of the Brown Dwarf Binary Separation Distribution ?}



\author{Viki Joergens}{
  address={Max-Planck Institut f\"ur Astronomie, 
     K\"onigstuhl~17, D-69117 Heidelberg, Germany}
}



\begin{abstract}
Searches for companions of brown dwarfs 
     by direct imaging probe mainly orbital separations 
     greater than 3--10\,AU. On the other hand, 
     previous radial velocity surveys of brown dwarfs are mainly sensitive to 
     separations smaller than 0.6\,AU.
     It has been speculated if the peak of the separation distribution of brown dwarf
     binaries lies right in the unprobed range.
Very recent work
for the first time extends high-precision radial velocity surveys of
     brown dwarfs out to 3\,AU (Joergens 2008).
Based on more than six years UVES/VLT spectroscopy 
     the binary frequency of brown dwarfs and (very) low-mass stars
     (M4.25-M8) in Chamaeleon\,I was determined:
it is 18$^{+20}_{-12}$\,\% for the whole sample 
     and 10$^{+18}_{-8}$\,\% for the subsample of ten brown dwarfs and very low-mass stars 
     (M$\leq 0.1\,M_{\odot}$).
     Two spectroscopic binaries were confirmed, these are
     the brown dwarf candidate ChaH$\alpha$\,8, 
     and the low-mass star CHXR\,74. 
     Since their orbital separations appear to be 1\,AU or greater,
     the binary frequency at $<$1\,AU might be less than 10\%.
Now for the first time companion searches of (young) brown dwarfs cover the 
     whole orbital separation range and the following observational constraints for 
     models of brown dwarf formation can be derived:
     (i) the frequency of brown dwarf and very low-mass stellar binaries at $<$3\,AU 
     is not significantly exceeding that at $>$3\,AU;
     i.e. direct imaging surveys do not miss a significant fraction of brown dwarf binaries;
     (ii) the overall binary frequency of brown dwarfs and very low-mass
     stars is 10-30\,\%;
     (iii) the decline of the separation distribution of brown dwarfs 
     towards smaller separations seem to occur between 1 and 3\,AU;
     (iv) the observed continuous decrease of the binary frequency from the stellar 
     to the substellar regime is confirmed at $<$3\,AU 
     providing further evidence for
     a continuous formation mechanism from 
     low-mass stars to brown dwarfs.

\end{abstract}

\maketitle


\section{Introduction}

Search for companions to brown dwarfs (BDs)
are of primary interest for understanding BD formation, for which
no widely accepted model exists 
(e.g. Luhman et al. 2007).
The frequency and properties of BDs in multiple systems are fundamental parameters
in formation models, e.g. embryo-ejection scenarios for BD formation
predict only few BD binaries in preferentially close orbits, while 
isolated fragmentation scenarios are expected to generate BD binaries with similar
(scaled-down) properties as stellar binaries have.
However, binary properties of BDs are poorly constrained for close separations: 
Most of the current surveys for companions
to BDs and very low-mass stars (VLMS, $M \leq 0.1\,M_{\odot}$)
are done by direct (adaptive optics or HST) imaging 
(e.g. Burgasser et al. 2007)
and are not sensitive to close separations
($a\leq$3\,AU).
These surveys find a lower frequency (10-30\%) of 
BD/VLM binaries compared to stars and
a separation distribution with a peak around 3--10\,AU.
The observed peak of the separation distribution is close to the incompleteness limit.
It has been suggested that BD/VLM binaries with a separation $<$3\,AU
are as frequent or even more frequent than those at $>$3\,AU 
(Pinfield et al 2003; Maxted \& Jeffries 2005; Burgasser et al. 2007).
Hence,
the peak of the BD/VLM binary separation distribution could lie below 3\,AU.
Given these limits of the current observational data, 
the question arises if our current picture of BD/VLM binaries is complete,
or, whether we miss a significant fraction of very close and/or
small mass ratio systems.

Spectroscopic monitoring for radial velocity (RV) variations provides a means to detect 
very close binaries. 
While several spectroscopic surveys for companions of 
BD/VLMS in young cluster
(e.g. Joergens 2006; Maxted et al. 2008)
and in the field
(e.g. Guenther \& Wuchterl 2003; Basri \& Reiners 2006)
have been started in recent years, data sampling is sparse in most cases and 
the number of confirmed close companions to BD/VLMS is still small
(only four, e.g. Stassun et al. 2006; Joergens \& M\"uller 2007).
Furthermore, the determination of the binary frequency of BD/VLMS 
in these surveys is limited to 
$\leq$0.1--0.6\,AU. 
This leaves still a significant unprobed gap between the separations studied by RV surveys and those
probed by direct imaging surveys ($>$3\,AU).
This was closed recently by follow-up UVES observations 
within the framework of an RV survey
in the Chamaeleon\,I star-forming region (Joergens 2008).
It allowed
for the first time to probe the important separation range 1--3\,AU for 
BD/VLMS, as outlined in the following.

\section{New results of RV survey in Cha\,I}

\begin{figure}
  \includegraphics[height=.3\textheight]{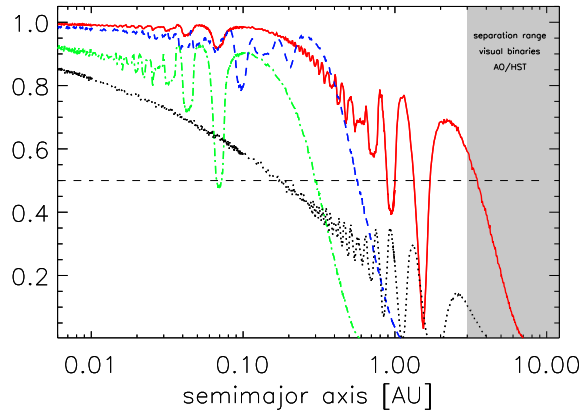}
  \caption{\label{fig:Pdetect}
Comparison of detection probability of different BD/VLMS radial velocity surveys.
Solid (red) line: Joergens 2008; 
dotted (black) line: Basri \& Reiners 2006; 
dashed (blue) line: Guenther \& Wuchterl 2003; 
dash-dotted (green) line: Kurosawa et al. 2006. 
Calculation based on Monte-Carlo simulation.
See text for more details on used parameters. From Joergens (2008).
}
\end{figure}

An RV survey is carried out for close (planetary and BD) companions
to very young BD/VLMS in Cha\,I with the UVES spectrograph at the VLT.
Including very recent data (Joergens 2008), this survey is sensitive to companions 
at orbital distances of 3\,AU and smaller with a
detection rate of 50\% or more, as shown by a
Monte-Carlo simulation (see Fig.\,\ref{fig:Pdetect}). 
For BD/VLM binaries with a mass ratio close to unity ($q\equiv M_2/M_1>$0.8),
the survey is sensitive to even larger separations.
This is a significant extension to larger orbital separations
compared to previous results of this survey
(Joergens 2006) as well as compared to other RV surveys of young BD/VLMS 
(Kurosawa et al 2006; Prato 2007; Maxted et al. 2008), which cover 
separations $\leq$0.3\,AU, and to other surveys of field BD/VLMS
(Reid et al 2002; Guenther \& Wuchterl 2003; Basri \& Reiners 2006), 
which cover separations $\leq$0.6\,AU 
(comparison based on $q$ between 0.2 and 1, 3.3\,$\sigma$ detection,
50\% detection probability; cf. Joergens 2008 for more details).
Thus, for the first time, the binary frequency of BD/VLMS 
is probed for the whole separation range $<$3\,AU with a sensitivity also to small mass ratio systems.

\begin{figure}[t]
\centering
\includegraphics[width=0.44\textwidth,angle=0,clip]{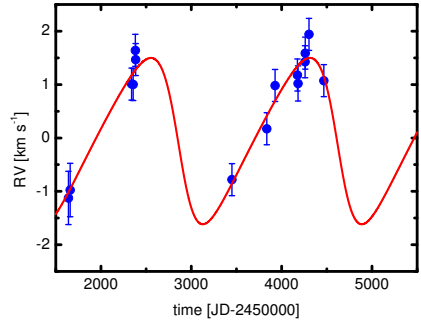}
\caption{
\label{cha8}
Discovery of low-mass RV companion orbiting the young BD/VLMS ChaH$\alpha$\,8 based on 
UVES/VLT data obtained between 2000 and 2008
(partly published in Joergens\,\&\,M\"uller\,2007).
Overplotted is the best-fit Kepler orbit
(semi-amplitude 1.6\,km\,s$^{-1}$, period 4.8\,years, eccentricity $e$=0.28).
The derived estimate (model dependent)
for the companion minimum mass M$_2\sin i$ is 17--22\,M$_{Jup}$.
}
\end{figure}

Two objects 
were identified as spectroscopic binaries based on significant RV variability
($\chi^2$ test with 3.3\,$\sigma$ threshold).
This corresponds to an observed binary fraction of
18$^{+20}_{-12}$\,\% for the whole sample of eleven BDs and 
(very) low-mass stars (M$\leq 0.25\,M_{\odot}$)
and 10$^{+18}_{-8}$\,\% for the subsample of ten BD/VLMS 
(M$\leq 0.1\,M_{\odot}$), respectively.
The detected binaries are:
(i) The BD/VLMS ChaH$\alpha$\,8 (M5.75-M6.5), which
has a substellar RV companion
in a $\sim$1\,AU orbit, as previously discovered
(Joergens \& M\"uller 2007, see Fig.\,\ref{cha8}). ChaH$\alpha$\,8
is likely the lowest mass RV companion found so far
around a BD/VLMS.
(ii) The low-mass star CHXR\,74 (M4.25-M4.5), which has a spectroscopic companion
in a relatively long period orbit (presumably $>$12~years).

These spectroscopic binaries appear to have 
orbital periods of at least a few years, i.e.
orbital separations of 1\,AU or greater.
There were no signs found for the presence of 
shorter period companions around the targets.
This is noteworthy because those cause 
larger RV signals and are, therefore,
easier to detect.
Thus, while the rate of BD/VLM binaries at $\leq$3\,AU is found
to be 10\% (18\% including CHXR\,74),
at separations $<$1\,AU it might be smaller than 10\%
(0/11 for the whole sample and 0/10 for the BD/VLMS subsample, respectively).

\section{Conclusions}

For the first time the binary frequency of BD/VLMS is probed
in the orbital separation range 1--3\,AU (Joergens 2008).
The separation distribution of BD/VLM binaries detected by direct imaging surveys
(e.g. Burgasser et al. 2007)
has a peak around 3--10\,AU which is close to the incompleteness limit of these surveys.
It seemed, therefore, possible that the largest fraction of BD/VLM binaries has
separations in the range of about 1--3\,AU and remained yet undetected.
The here determined BD/VLM binary frequency of 
10$^{+18}_{-8}$\,\%
at separations
of 3\,AU and smaller shows that this is not the case: the rate of 
spectroscopic binaries at $<$3\,AU is not significantly exceeding 
the rate of resolved binaries at larger separations 
(10--30\%, e.g. Burgasser et al. 2007).
Thus, the separation range $<$3\,AU missed by direct
imaging surveys is not adding a significant fraction to the BD/VLM binary frequency
and the overall binary frequency of BD/VLMS is in the range 10-30\%.
The finding of no signs (0/10, i.e. $\leq$10\%) for companions at $<$1\,AU
is consistent with a decline of the separation distribution of BD/VLM binaries
towards smaller separations starting between 1 and 3\,AU.
Further, the previous finding that the observed mass-dependent decrease of the stellar binary frequency
extends continuously into the BD regime (e.g. Burgasser et al. 2007)
is confirmed here for separations $<$3\,AU.
This is consistent with a continuous formation mechanism from 
low-mass stars to BDs.
 
The here presented results are based on a relatively small sample.
By combining this work in Cha\,I with surveys of other young regions,
which are restricted to separations $\leq$0.3\,AU,
an overall observed binary fraction of very young 
BD/VLMS can be determined with better statistics but with a more limited
separation range. In this way, a binary frequency of 7$^{+5}_{-3}$\% (7/97) is found
at $\leq$0.3\,AU.
In the near future, the sample of the survey in Cha\,I will be substantially 
enlarged allowing the re-investigation of the 
binary frequency at $<$3\,AU of BD/VLMS in Cha\,I with significantly improved statistics
for a uniformly obtained data set. 
Current and future observational efforts are, in addition, directed towards
determination of orbital parameters of the detected binary systems 
including RV follow-up and high-resolution imaging/astrometry.

\end{document}